\documentclass[10pt, conference,letterpaper]{IEEEtran}
\IEEEoverridecommandlockouts
\usepackage{cite}
\usepackage{amsmath,amssymb,amsfonts}
\usepackage{algorithmic}
\usepackage{graphicx}
\usepackage{textcomp}
\usepackage{url}
\usepackage{xcolor}

\DeclareMathOperator*{\argmin}{arg\,min}
\def\BibTeX{{\rm B\kern-.05em{\sc i\kern-.025em b}\kern-.08em
    T\kern-.1667em\lower.7ex\hbox{E}\kern-.125emX}}
\begin{document}

\title{Decentralized Online Federated G-Network Learning for Lightweight Intrusion Detection\thanks{This research has been supported by the European Commission Horizon Europe – the Framework Programme for Research and Innovation (2021-2027) DOSS Project under Grant Agreement No: 101120270.}
}

\author{\IEEEauthorblockN{Mert Nak\i p}
\IEEEauthorblockA{\textit{Inst. Theoretical \& Appl. Informatics} \\
\textit{Polish Academy of Sciences (IITIS-PAN)}\\
Gliwice, Poland \\
mnakip@iitis.pl}
\and
\IEEEauthorblockN{Baran Can G\"{u}l}
\IEEEauthorblockA{\textit{Inst. of Industrial Automation and} \\ \textit{Software Engineering} \\
\textit{University of Stuttgart}\\
Stuttgart, Germany \\
baran-can.guel@ias.uni-stuttgart.de}
\and
\IEEEauthorblockN{Erol Gelenbe}
\IEEEauthorblockA{\textit{Inst. Theoretical \& Appl. Informatics} \\
\textit{Polish Academy of Sciences (ITIS-PAN)}\\
Gliwice, Poland\\
\textit{Universit\'{e} C\^{o}te d'Azur, CNRS, I3S} \\
06100 Nice, France \\
\textit{Ya\c{s}ar University}
Izmir, T\"{u}rkiye \\
seg@iitis.pl}
}

\maketitle

\begin{abstract}
Cyberattacks are increasingly threatening networked systems, often with the emergence of new types of unknown (zero-day) attacks and the rise of vulnerable devices. Such attacks can also target multiple components of a Supply Chain, which can be protected via  Machine Learning (ML)-based Intrusion Detection Systems (IDSs). However, the need to learn large amounts of labelled data often limits the applicability of ML-based IDSs to cybersystems that only have access to private local data, while distributed systems such as Supply Chains have multiple components, each of which must preserve its private data while being targeted by the same attack To address this issue, this paper proposes a novel Decentralized and Online Federated Learning Intrusion Detection (DOF-ID) architecture based on the G-Network model with collaborative learning, that allows each IDS used by a specific component to learn from the experience gained in other components,  in addition to its own local data, without violating the data privacy of other components. The performance evaluation results using public Kitsune and Bot-IoT datasets show that DOF-ID significantly improves the intrusion detection performance in all of the collaborating components, with acceptable computation time for online learning. 
\end{abstract}


\section{Introduction}

Much work has been done over the years to secure cyber-physical systems \cite{Caglayan1,Caglayan2,Gelenbe2,Skarmeta,Caglayan3}. Intrusion Detection  Systems (IDS), are important components of such overall cybersecurity schemes which have 
often been developed using Machine Learning to detect anomalies and threats
in incoming network traffic \cite{alsoufi2021anomaly,maniriho2020anomaly,CDIS}, including multiclass classification to detect different types of attacks in a unified manner \cite{nisioti2018intrusion,sarker2021cyberlearning}.

Federated Learning (FL), also known as collaborative learning, is a machine learning technique that trains a machine learning algorithm via multiple independent sessions, each using its own dataset, presenting several advantages and novel research issues\cite{Kairouz}. This approach contrasts with traditional centralized machine learning techniques where local datasets are merged into one training session, as well as with approaches that assume that local data samples are statistically identical. Thus it is particularly useful when multiple entities such as Supply Chains \cite{Supply1,Supply2}, emergency evacuation systems \cite{Gelenbe2}, autonomous vehicles \cite{Frotscher}, or systems where different types of IoT devices concurrently operate \cite{Pokhrel,Savazzi}, and when heterogeneous active objects collaborate in a common physical and data driven environment \cite{Xu}.

For instance, multiple components of a Supply Chain may experience partially distinct and partially similar types of attacks \cite{Supply3,Supply4,Supply5}, at different traffic rates and different attack frequencies. Also, the types of attacks and frequency of attacks experienced by a given entity (e.g. a specific sub-system) may be closely connected to its own physical environment which is shared among several flows of supplies, and have different levels of cost, delay and profitability, which that particular a particular sub-system may not wish to share with other systems for competitive reasons. In such cases a ``concurrent'' yet ``federated'' type of learning can be used to design an appropriate IDS that may be useful to all sub-systems, provided they do not directly access each other's data.

In recent work G-Networks \cite{G-Networks93}, which are a generalization of the Random Neural Network \cite{RNN} and of ``queueing networks with negative and positive customers'', have been successfully used for cyberattack detection and IDS \cite{Brun} with deep learning. Furthermore, in \cite{G_networks} it was shown that appropriately designed auto-associative G-Network models can very accurately detect multiple types of attacks simultaneously with training that is only based on ``benign'' traffic. 
Thus, in this paper we extend this approach using G-Networks for attack detection to Federated Learning where the mix of multiple attacks may vary between distinct sites that share their learning experience but do not share their private data.

\subsection{Related Work on Federated Learning and IDS}

FL-based IDS is typically centralized or decentralized FL. The former collects updates of the learning algorithm in a central server so as to build a global model, while in decentralized FL, training is performed locally by each separate concurrent training site and the updated algorithms are then transferred among the separate sites. 

\subsubsection{Centralized Federated Learning}
In \cite{Taheri_2020} a centralized federated architecture is developed to detect malware using the Generative Adversarial Network (GAN) for the industrial Internet of Things (IoT). Although this architecture achieves high accuracy in detecting attacks, it assumes that a validation set is available at the server, which may partially violate privacy of the user data. In \cite{Li_Wu_Jiang_2022} a multi-step Distributed Denial of Service (DDoS) attack prediction method that uses Hidden Markov Model within a centralized FL architecture using Reinforcement Learning is presented and tested against a global algorithm, while  in \cite{Campos_2022} a multi-class classifier for FL-based IDS considers multiple data distributions, and in \cite{Nguyen_2019}, a self-learning distributed approach is developed to detect IoT devices compromised by Mirai malware. Similarly, \cite{Mothukuri_2021} presents an anomaly detection approach based on centralized FL to classify and identify attacks in IoT networks, and has tested it on a dataset consisting of Man-in-the-Middle (MitM) and flood attacks. In \cite{Li_2021}, an architecture was proposed to mitigate DDoS in industrial IoT networks offering reduced mitigation delay. 


\subsubsection{Decentralized Federated Learning}
In order to defend only against gradient attacks, Reference \cite{Lu_2023} proposed a decentralized FL framework, which is based on a peer-to-peer network for sending, aggregating, and updating local models. 
Another study \cite{Lian_2022} used decentralized FL to detect anomalies in network traffic generated by IoT devices, when all federated IDSs are shared with each distinct participant to obtain a weighted average, while
in \cite{Al_2022} blockchain-based FL was used as a decentralized architecture to specifically detect poisoning attacks. 

\begin{figure*}[t!]
	\centering
	\includegraphics[width=17cm, height=10cm]{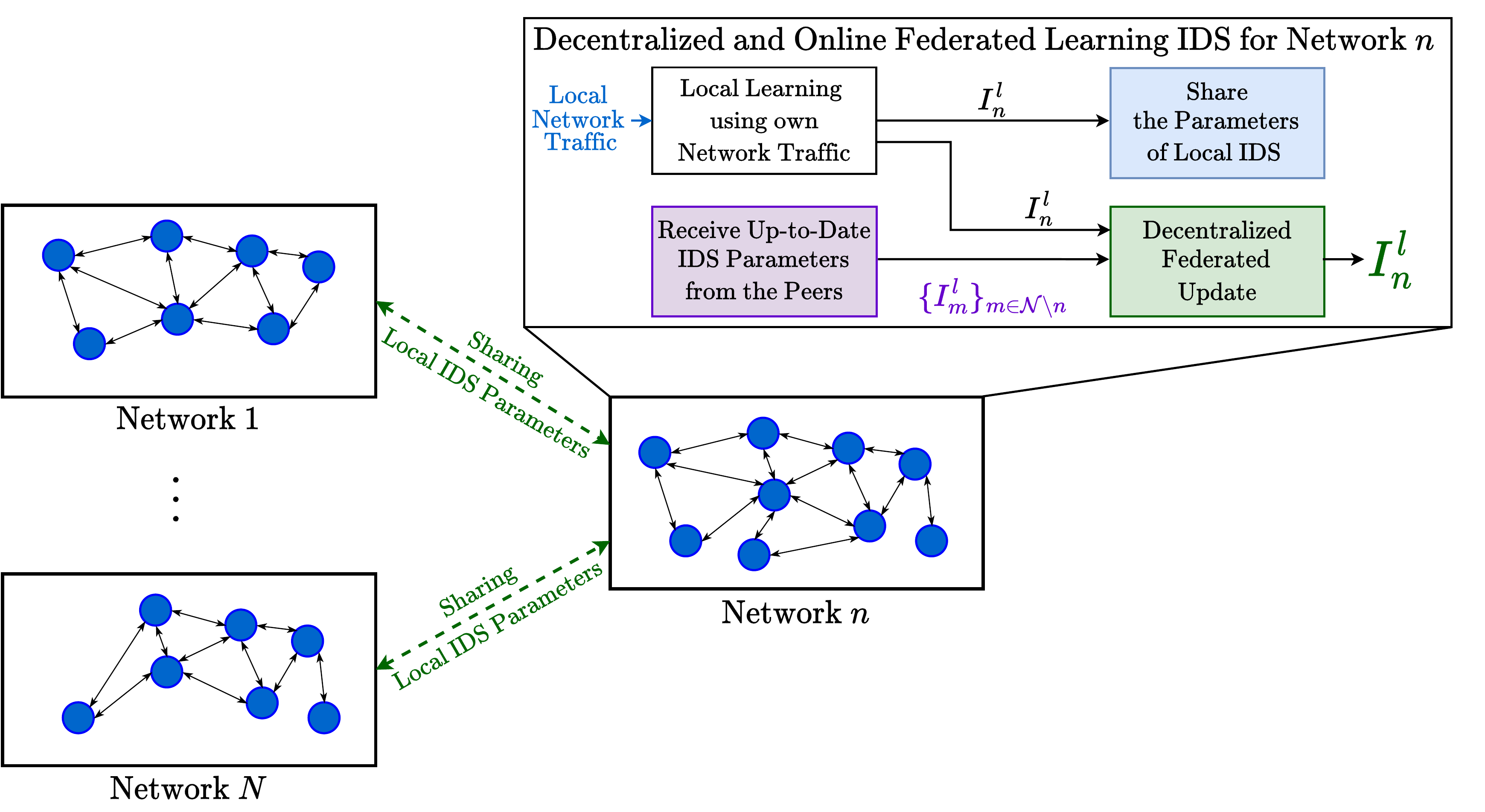}
	\caption{Schematic system representation of the Decentralized and Online Federated Learning Intrusion Detection (DOF-ID) architecture.}\label{fig:FL_IDS}
\end{figure*}

\subsection{Contributions of This Paper}

In this paper, we propose a novel Decentralized Online Federated Learning Intrusion Detection (DOF-ID) architecture for improved online learning of ML-based IDS, that uses the Deep Random Neural Network \cite{Deep1}. The DOF-ID architecture hosts many distinct components of a supply chain, each component having potentially several IoT or IP nodes, each of which utilizes an instance of a common IDS, learns directly from its local data, and collaborates with other components to incorporate their up-to-date knowledge into its IDS. This architecture improves the overall security of all collaborating nodes with online learning between nodes, by taking advantage of the experience of each node,  while preserving the confidentiality of the local data at each of  these nodes. 

The DOF-ID architecture uses a learning procedure that combines Local Learning, and Decentralized Federated Updates (DFU) with concurrent parameter updates taking place on the collaborating nodes with local data. Therefore the proposed DOF-ID architecture with the DFU algorithm contrasts sharply with recent work on federated learning IDS.


We then evaluate the performance of the DOF-ID architecture and compare it with the performance of four benchmark methods, on three different types of cyberattacks obtained from two well-known public datasets: Kitsune \cite{kitsune_paper, kitsune_dataset} and Bot-IoT \cite{botiot_dataset}. 

The results show that DOF-ID provides significant performance gains compared to learning from local data alone and outperforms other state-of-the-art federated learning methods.

The remainder of this paper is organized as follows: Section~\ref{sec:Method} presents the novel DOF-ID architecture with DRNN-based IDS (Section~\ref{sec:IDS}), local learning algorithm (Section~\ref{sec:Local_Learning}), and the DFU algorithm (Section~\ref{sec:Federated_Learning}). Section~\ref{sec:Results} evaluates the performance of the proposed DOF-ID architecture on public datasets. Section~\ref{sec:Conclusion} summarizes the paper and presents some insights towards future work.

\section{Intrusion Detection with Decentralized and Online Federated Learning}\label{sec:Method}

In order to improve the performance of an ML-based IDS, we now present a novel Decentralized and Online Federated Learning Intrusion Detection Intrusion Detection (DOF-ID) architecture, which is based on the collaboration of $N$ nodes (denoted by set $\mathcal{N}$) using separate local instances of the same IDS. Figure~\ref{fig:FL_IDS} displays the proposed architecture from the perspective of a particular node $n$, where the nodes are represented as computer networks (e.g. Internet of Things (IoT) networks). From the application perspective, one may consider DOF-ID as a subscription based service, where each subscriber (i.e. node) receives the updates of other subscribers to improve its local security level. 


As seen in Figure~\ref{fig:FL_IDS}, each node $n$ directly communicates with other nodes in $\mathcal{N}$ (i.e. peers) to send locally learned parameters of IDS and to receive those learned by other nodes. That is, locally learned IDS parameters are Peer-to-Peer (P2P) shared between every node in the DOF-ID architecture distributing the knowledge over collaborating nodes in $\mathcal{N}$ to improve their security (subsequently the global security) while the confidentiality of local data in every node is assured. 

The DOF-ID architecture operates over time windows each with a length of $T$ seconds, where time windows are considered to be synchronized among collaborating nodes. We also assume that the first time window (denoted by $l = 0$) starts with the use of DOF-ID architecture. Accordingly, at the beginning of each time window $l$, each node $n$ updates its local IDS if no intrusion is detected in the previous window $l-1$. That is, if the intrusion decision of node $n$ in window $l-1$, denoted by $y_n^{l-1} \in \{0, 1\}$, equals zero, node $n$ executes the following steps as part of the learning procedure for the current window $l$: 
\begin{enumerate}
	\item It learns from training data containing local \emph{benign} network traffic for windows up to beginning of $l$, denoted by $\mathcal{D}_n^l$, such that $\mathcal{D}_n^l = \{\, k: ~~ y_n^k = 0, \, \forall k \in \{1, \dots, l-1\} \, \}$. When the learning is completed, an up-to-date locally trained IDS of node $n$, denoted by $I_n^l$, is obtained to use for detection in window $l$.
	\item It shares the parameters of $I_n^l$ with other collaborating nodes in $\mathcal{N}$ and receives the local updates of those nodes, i.e. $\{I_{n'}^l\}_{n' \in \mathcal{N}\setminus n}$. In this paper, it is assumed that the P2P parameter exchange is instantaneous; however, future work shall analyse the time, bandwidth and energy requirements of the proposed DOF-ID architecture regarding P2P parameter exchange.
	\item As the final step, node $n$ updates the local IDS $I_n^l$ by merging its parameters with $\{I_{n'}^l\}_{n' \in \mathcal{N}\setminus n}$ via the proposed DFU.
\end{enumerate} 
Following the training procedure in window $l$, each node $n$ estimates the intrusion probability $y_n^l$ through the following steps: 
\begin{enumerate}
    \setcounter{enumi}{4}
    \item The inputs of the utilized IDS are considered to be statistics calculated from the traffic of node $n$. Thus, node $n$ first calculates traffic statistics as a vector of IDS inputs, denoted by $x_n^l$, in time window $l$.
    \item Using the up-to-date IDS $I_n^l$, the final intrusion decision $y_n^{l} \equiv I_n^l(x_n^l)$ for traffic statistics $x_n^l$ is calculated. 
\end{enumerate}

In the rest of this section, we respectively present our methodology for the particular ML-based IDS utilized in DOF-ID architecture as well as the local and federated learning algorithms.

\subsection{IDS Utilized in the DOF-ID Architecture}\label{sec:IDS}

\begin{figure}[h!]
	\hspace{-0.75cm}
	\includegraphics[scale=0.65,keepaspectratio]{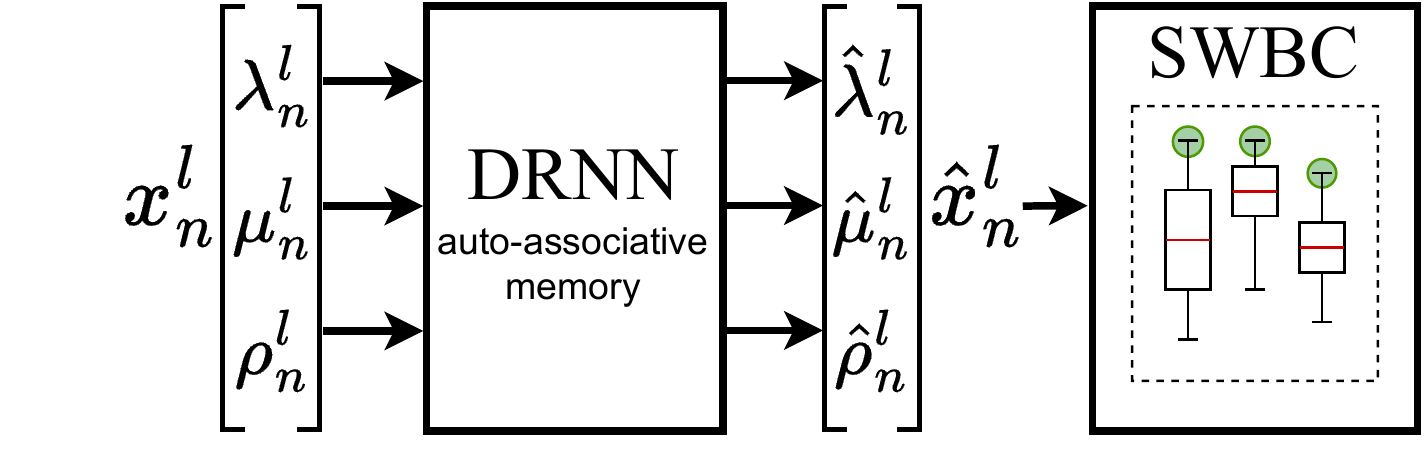}
	\caption{Structure of the IDS utilized in the DOF-ID architecture}\label{fig:IDS}
\end{figure}

Within our DOF-ID architecture, we use an IDS which is the modified version of the one presented in \cite{G_networks} and comprised of DRNN and Statistical Whisker-based Benign Classifier (SWBC) as shown in Figure~\ref{fig:IDS}. At each window $l$, this IDS estimates $y_n^l$ that indicates whether the traffic of the considered node $n$ in window $l$ is malicious based on the input vector of traffic statistics, $x_n^l$.


\subsubsection{Traffic Statistics}

Let $p_n^t$ denote the packet with length $|p_n^t|$ generated in node $n$ at instantaneous time $t$, and let $P_n^l$ be the set of all packets generated in $n$ within time window $l$ whose length equals $T_n$:
\begin{equation}
	P_n^l = \{\,  p_n^t: ~(l-1) \, T_n \leq t < l \, T_n \, \}.
\end{equation} 

In each time window $l$, node $n$ calculates three main statistics that represent the overall density of the network traffic as the average packet length in bytes ($\mu_n^l$), the average number of packets per second ($\lambda_n^l$), and the average traffic in bytes per second ($\rho_n^l$):
\begin{equation}
	\mu_n^l = \frac{\sum_{p \in P_n^l} |p|}{|P_n^l|}, \quad  \lambda_n^l = \frac{|P_n^l|}{T_n}, \quad \rho_n^l = \frac{\sum_{p \in P_n^l} |p|}{T_n}.
\end{equation} 
In order to use these statistics with DRNN, each element $i$ of $x_n^l = [\mu_n^l, \lambda_n^l, \rho_n^l]$, denoted by $x_{n, i}^l$, is normalized to have values in $[0, 1]$. 

\subsubsection{Deep Random Neural Network to Create Auto-Associative Memory}
In order to create an auto-associative memory, we use the well-known lightweight deep learning model DRNN \cite{Deep1}, which is a Random Neural Network \cite{RNN} model with feed-forward and clustered structure. As a result of its unique architecture presented in \cite{Deep1}, each neuron at hidden layers of DRNN utilizes the following activation function, which is specific to this model: 
\begin{eqnarray}\label{eq:activation}
	\Psi(\Lambda) &=& \frac{p \, (r + \lambda^+) + \lambda^- + \Lambda }{ 2 \, [\lambda^- + \Lambda ]}\\
	& - &\sqrt{ {\Big( \frac{p \, (r + \lambda^+) + \lambda^- + \Lambda }{ 2 \, [\lambda^- + \Lambda ]} \Big)}^2 - \frac{\lambda^+}{\lambda^- + \Lambda }}~, \nonumber
\end{eqnarray}
where $\Lambda$ is the input of the given cluster, $p$ is the probability that any neuron received trigger transmits a trigger to some other neuron, and $\lambda^+$ and $\lambda^-$ are respectively the rates of external Poisson flows of excitatory and inhibitory input spikes to any neuron. On the other hand, the neurons at the output layer of DRNN utilize linear activation functions.

As we consider three different network statistics, the DRNN model that we use in this paper consists of $H=3$ fully connected layers with three neurons each. Accordingly, from the input vector $x_n^l$, DRNN estimates vector $\hat{x}_n^l$ of the statistics that are expected to be observed when the network traffic is benign: 
\begin{align}
	&\hat{x}_{(n, 1)}^l = \Psi([x_n^l, 1] \, W_{(n, 1)}^l)\label{eq:forward1_DRNN}\\
	&\hat{x}_{(n, h)}^l =  \Psi([\hat{x}_{(n, h-1)}, 1] \, W_{(n, h)}^l) ~~ \forall h \in \{2, \dots, H-1\},\label{eq:forward_DRNN}\\
	&\hat{x}_n^l =  [\hat{x}_{(n, h-1)}^l, 1] \, W_{(n, H)}^l,\label{eq:output_DRNN}
\end{align}
where $\hat{x}_{(n, h)}^l$ is the output of layer $h$, and $\hat{x}_n^l$ is the final output of DRNN for node $n$ in window $l$. In addition, the term $[x_n^l, 1]$ or $[\hat{x}_n^l, 1]$ indicates that $1$ is added to the input of each layer as a multiplier of the bias, and $W_{(n, h)}^l$ is the connection weight matrix between layers $h-1$ and $h$ of DRNN in $I_n^l$.

\subsubsection{Statistical Whisker-based Benign Classifier}
As the second operation in $I_n^l$ (IDS of node $n$ in time window $l$) that makes a decision on an intrusion, SWBC is used to measure the significance of the difference between the actual statistics measured from the network traffic and the expected statistics estimated by DRNN. SWBC is originally proposed in \cite{G_networks} and calculates the decision $y_n^l$ as follows: 
\begin{equation}
	\zeta_n^l = \sum_{i \in \{1, 2, 3\}} \mathbf{1}(|x_{n, i}^l - \hat{x}_{n, i}^l| > w_{n, i}^l),
\end{equation}
\begin{equation}
	y_n^l  = \mathbf{1}(\zeta_n^l > \theta_n^l),
\end{equation}
where $x_{n, i}^l$ is the $i$--th element of vector $x_n^l$ corresponding the traffic statistic $i$, and $\{w_{n, i}^l\}_{i \in \{1, 2, 3\}}$ and $\theta_n^l$ are the only parameters of the decision maker which are computed (learned) during training along with the connection weights of DRNN.

\subsection{Local Learning}\label{sec:Local_Learning}

We now present the methodology of the local learning procedure that node $n$ executes to learn parameters of $I_n^l$ only using local data. In this procedure, node $n$ respectively learns the DRNN weights and SWBC parameters for window $l$ based on the available data $\mathcal{D}_n^l$.  

\subsubsection{Learning DRNN Weights}

Using the local data of node $n$, DRNN in $I_n^l$ is trained to create an auto-associative memory for the normal -- benign -- network traffic. To this end, first the connection weights of each hidden layer $h \in \{ 1, \dots, H-1\}$ are calculated by minimizing a square cost with L1 regularization via Fast Iterative Shrinkage-Thresholding Algorithm (FISTA) with the following objective: 
\begin{align}\label{eq:FISTA_objective}
	W_{(n ,h)}^l & = \arg\hspace{-0.5cm}\min_{\{W: \, W\geq \, 0\}} \, \Bigl(  \\
	& \, \Bigl| \Bigl| [adj(\Psi(\hat{X}_{(n, h-1)}^l W_{R~})), \mathbf{1}_{|\mathcal{D}_n^l|}] \, W - \hat{X}_{(n, h-1)}^l \Bigr| \Bigr|_{2}^2 \nonumber \\
	&+ ||W||_{1} \, \Bigr)~,\nonumber
\end{align}
where $\hat{X}_{(n, h-1)}^l$ is the matrix of $\hat{x}_{(n, h-1)}^k$ collected for $k \in \mathcal{D}_n^l$ for $h \geq 1$, $\hat{X}_{(n, 0)}^l = X_n^l$ which is the matrix of $x_n^k$ collected for $k \in \mathcal{D}_n^l$, $\mathbf{1}_{|\mathcal{D}_n^l|}$ is a column vector of ones with length $|\mathcal{D}_n^l|$, and $W_{R~}$ is randomly generated ($H \times H$) matrix with elements in the range $[0, 1]$. In addition, $adj(A)$ linearly maps the elements of matrix $A$ to the range $[0, 1]$ then applies z-score, and adds a positive constant to remove negativity. 

For each layer $h \in \{1, \dots, H-1\}$, after FISTA is executed, we normalize each resulting weight matrix $W_{(n ,h)}^l$: 
\begin{equation}
	W_{(n ,h)}^l \leftarrow 0.1\frac{W_{(n ,h)}^l}{\max_{k \in \mathcal{D}_n^l}\big(\hat{X}_{(n, h)}^k)}~.
\end{equation}
The connection weights of the output layer $H$ are calculated via an extreme learning machine as
\begin{equation}\label{eq:ELM}
	W_{(n, H)}^l =  (\hat{X}_{(n, H-1)}^l)^+ X_n^l,
\end{equation}
where $A^+$ denotes the pseudo-inverse of matrix $A$.

\subsubsection{Computing SWBC Parameters}


Using the training data, $\mathcal{D}_n^l$, which consists of only benign traffic features, we determine the values of $\theta_n^l$ and $w_{n, i}^l$ for each statistic $i$. To this end, for each $i$, the value of the absolute difference $z_{n, i}^k = |x_{n, i}^k - \hat{x}_{n, i}^k|$ is computed for all $k\in\mathcal{D}_n^l$. 

Then, we compute the lower quartile $Q_{n, i}^L$ and upper quartile $Q_{n, i}^U$ of $\{z_{n, i}^{k}\}_{k\in\mathcal{D}_n^l}$. 
Using $Q_{n, i}^L$ and $Q_{n, i}^U$, the upper whisker $w_{n, i}^l$ is calculated as 
\begin{equation}
	w_{n, i}^l = Q_{n, i}^U + \frac{3}{2}(Q_{n, i}^U - Q_{n, i}^L) \qquad \forall i\in \{1, 2, 3\}
\end{equation}

Since the training data contains only benign traffic, $\theta_n^l$ must be selected to classify training samples as benign traffic. Meanwhile, we should also consider that the training data may include false negative samples. Therefore, we determine $\theta_n^l$ to classify the majority but not all of the training samples as benign traffic, and we set the value of $\theta_n^l$ to the mean of $\zeta_n^l$ (i.e. the average number of abnormal statistics) plus two standard deviations of $\zeta_n^l$ in $\mathcal{D}_n^l$:
\begin{equation}
	\theta_n^l = \text{mean}_{\mathcal{D}_n^l}(\zeta_n^l) + 2\, \text{std}_{\mathcal{D}_n^l}(\zeta_n^l)
\end{equation}

\subsection{Decentralized Federated Update}\label{sec:Federated_Learning}

We now present the Decentralized Federated Update (DFU) algorithm that is performed as the last step of our DOF-ID architecture. In the DFU algorithm, the parameters of node $n$ are updated using the parameters of other nodes in DOF-ID, whose data is unknown by node $n$. To this end, at each window $l$ in this algorithm, node $n$ performs three main operations: 1) select the set of concurring nodes, denoted by $\mathcal{C}_n^l$ that achieve decisions similar to those of node $n$, 2) update the value of each parameter segment in $I_n^l$ using the corresponding segment with closest value to that segment among all nodes in $\mathcal{C}_n^l$, 3) recalculate the output layer weights of DRNN via extreme learning machine in order to fully adapt updated parameters to the local network traffic. 

\subsubsection{Selecting a Set of Concurring Nodes}

In the current window $l$, node $n$ first selects a set of nodes that concur with it for most of its decisions regarding local data. In order to select the concurring nodes, node $n$ evaluate the performance of each node $m \in \mathcal{N}\setminus n$ on the local data of node $n$ over all time windows up to current window $l$: 
\begin{equation}
    \mathcal{C}_n^l = \{m :~~\frac{1}{l}\sum_{k=1}^{l} \mathbf{1}\bigl( I_m^l(x_n^k) = y_n^k \bigr) \geq \Theta, ~~ \forall m \in \mathcal{N}\setminus n \}
\end{equation}

\subsubsection{Updating IDS Parameters}

Using the IDSs of the concurring nodes, the parameters of $I_n^l$ are updated separately for each segment of the IDS (such as each DRNN layer, each SWBC whisker, and the SWBC threshold) averaging with the closest one for that segment among all the concurring nodes. 

To this end, first, for each layer $h \in \{1, \dots, H-1\}$ of DRNN in $I_n^l$, the node $m^*_h$ that has the closest connection weights $W_{(n, h)}^l$ with node $n$ in window $l$ is obtained:
\begin{equation}\label{eq:closest_layer}
m^*_h = \argmin_{m \in \mathcal{C}_n^l}{\Biggl( \, \Bigl|\Bigl| W_{(n, h)}^l - W_{(m, h)}^l \Bigr| \Bigr|_1 \, \Biggr)}.
\end{equation}
Then, the connection weights of this layer, $W_{(n, h)}^l$, are updated as 
\begin{equation}\label{eq:update_closest_layer}
W_{(n, h)}^l \leftarrow c \, W_{(n, h)}^l + (1-c) \, W_{(m^*_h, h)}^l, 
\end{equation}
where $0.5 \leq c \leq 1$ is a coefficient of weighted averaging that prioritizes locally learned weights over federated weights. 

Similarly, for each whisker $w_{n, i}^l$ of SWBC in $I_n^l$, the node $m^*_i$ with the whisker value $w_{m^*_i, i}^l$ closest to $w_{n, i}^l$ is obtained among concurring nodes in window $l$:
\begin{equation}\label{eq:closest_whisker}
m^*_i = \argmin_{m \in \mathcal{C}_n^l}{\Biggl( \, \Bigl|w_{n, i}^l - w_{m, i}^l \Bigr| \, \Biggr)},
\end{equation}
and each whisker $w_{n, i}^l$ of SWBC in $I_n^l$ is updated: 
\begin{equation}\label{eq:update_closest_whisker}
w_{n, i}^l \leftarrow c\, w_{n, i}^l + (1-c) \, w_{m^*_i, i}^l .
\end{equation}
The decision threshold $\theta_n^l$ is also updated as
\begin{equation}\label{eq:update_closest_threshold}
\theta_n^l \leftarrow c \, \theta_n^l + (1-c) \, \theta_{m^*_\theta}^l
\end{equation}
for 
\begin{equation}\label{eq:closest_threshold}
    m^*_\theta = \argmin_{m \in \mathcal{C}_n^l}{\Biggl( \, \Bigl| \theta_n^l - \theta_m^l \Bigr| \, \Biggr)}.
\end{equation}

\subsubsection{Adapting the Updated IDS to Local Network Traffic}
Finally, the output layer weights of DRNN in the IDS $I_n^l$ are updated to fully adapt $I_n^l$ to local \emph{benign} network traffic of node $n$. To this end, (\ref{eq:ELM}) is repeated: 
\begin{equation}
	W_{(n, H)}^l =  (\hat{X}_{(n, H-1)}^l)^+ X_n^l.
\end{equation}

\section{Experimental Results}\label{sec:Results}
We now evaluate the performance of the proposed DOF-IDS architecture. To this end, from two publicly available datasets Kitsune \cite{kitsune_dataset} and Bot-IoT \cite{botiot_dataset}, we use three attack data each of which corresponds to a single node in DOF-IDS architecture. That is, we consider three collaborating nodes each of which is an IoT network whose data is obtained from a public dataset. 

We perform the experiments on a computer with 16 GB of ram and M1 Pro 8-core 3.2 GHz processor. The performance of DOF-IDS is also compared against four benchmark methods.

\subsection{IoT Traffic Datasets and Their Processing}

As the first node in DOF-IDS architecture, we use ``Mirai Botnet'' attack data from the Kitsune dataset \cite{kitsune_dataset}, which is the 
collection of $764,137$ individual traffic packets, which are transmitted by $107$ unique IP addresses within $7137$ seconds (approximately $2$ hours). 

As the second and third nodes in DOF-IDS, we use ``DoS HTTP'' and ``DDoS HTTP'' attacks from Bot-IoT dataset \cite{botiot_dataset}. The DoS HTTP attack data contains $29,762$ packets transmitted in $49$ minutes, and the DDoS HTTP attack data contains $19,826$ packets transmitted in $42$ minutes. Since these two types of attacks start this data with an attack traffic and the presented system requires cold-start with only benign traffic, we use each of these data by flipping it on the time axis.

During our experimental results in order to obtain approximately the same number of time windows from each dataset, we set the values of $T_n$ as follows: $23$ for Mirai, $9$ for DoS HTTP, and $8$ for DDoS HTTP. 
One should also note that both datasets include a binary ground truth $a(p_n^t)$ for each packet $p_n^t$, which is determined by the providers, stating whether the packet is a normal ``benign'' packet or ``malicious'' corresponding to an ongoing attack. Accordingly, based on the individual packet ground truths, we determine an overall ground truth, denoted by $g_n^l$, for each node $n$ in each time window $l$: 
\begin{equation}
    g_n^l = \mathbf{1}\Biggl(\frac{\sum_{p \in P_n^l}{a(p)}}{|P_n^l|} > 0.5 \Biggr)
\end{equation}

\subsection{Benchmark Methods}

\subsubsection{No Federated}
In this method, the contributions of the other nodes are not considered in the learning of the IDS parameters. This is the conventional training approach, which is equivalent to the local learning procedure of our DOF-IDS architecture. 

\subsubsection{Average over All Collaborating Nodes}
The rest of the benchmark methods are used in place of the DFU algorithm to update the IDS parameters after local learning.

In this method called ``Average'', the parameters of IDS in node $n$ (i.e. $I_n^l$) are updated as the average of all connection weights over all collaborating nodes in the DOF-ID architecture. To this end, connection weights for each layer $h$ of DRNN in $I_n^l$ are updated as 
\begin{equation}
W_{(n, h)}^l \leftarrow \frac{1}{N}\sum_{m\in \mathcal{N}}{W_{(m, h)}^l},
\end{equation}
Subsequently, SWBC parameters are also updated in the same way: 
\begin{equation}
w_{n, i}^l \leftarrow \frac{1}{N}\sum_{m\in \mathcal{N}}{w_{m, i}^l}~~ \forall i, ~~ \text{and} ~~  \theta_n^l \leftarrow \frac{1}{N}\sum_{m\in \mathcal{N}}{\theta_n^l}
\end{equation}
One should note that the parameters updated using this method become the same for all nodes. That is, at the end of this method, $I_n^l = I_m^l, ~ \forall n, m \in \mathcal{N}$.  

\subsubsection{Average with Closest Node}

In this method called ``Average with Closest Node (ACN)'', the parameters of $I_n^l$ are updated by taking their average with the closest parameters among all nodes in the DOF-ID architecture. To this end, first, the node $m^*$ that has the closest parameters with node $n$ at time window $l$ is obtained: 
\begin{align}
m^* = \argmin_{m \in \mathcal{N}\setminus n}\Biggl(&\sum_{h=1}^{H} \Bigr|\Bigr| W_{(n, h)}^l - W_{(m, h)}^l \Bigr| \Bigr|_1 \\
&+ \sum_{i=1}^{3} \Bigr| w_{n, i}^l -  w_{m, i}^l \Bigr|  \, + \,  \Bigr|\theta_n^l - \theta_m^l \Bigr| \, \Biggr) \nonumber
\end{align}
Then, in time window $l$, the parameters $I_n^l$ are updated taking their average with the parameters of $I_{m^*}^l$ as
\begin{align}
&W_{(n, h)}^l \leftarrow \frac{W_{(n, h)}^l + W_{(m^*, h)}^l}{2},~~ \forall h \\
&w_{n, i}^l \leftarrow \frac{w_{n, i}^l + w_{m^*, i}^l}{2},~~ \forall i \qquad
\theta_n^l \leftarrow \frac{\theta_n^l + \theta_{m^*}^l}{2} \nonumber
\end{align}

\subsubsection{Average with Closest Node per Layer}

In the last benchmark method called ``Average with Closest Node per Layer (ACN-L)'', the parameters of $I_n^l$ are updated for each parameter segment of the IDS (such as a layer of DRNN, a whisker of SWBC, and the threshold of SWBC) individually taking the average with the same part of the closest node. 

For each layer $h$ of DRNN in $I_n^l$, the connection weights of this layer are updated using (\ref{eq:update_closest_layer}) for a value of $m_h^*$ calculated using (\ref{eq:closest_layer}).
Then, each whisker $w_{n, i}^l$ of SWBC in $I_n^l$ is updated (\ref{eq:update_closest_whisker}) for a value of $m_i^*$ calculated using (\ref{eq:closest_whisker}). 
The decision threshold $\theta_n^l$ is updated by subsequently using (\ref{eq:closest_threshold}) and (\ref{eq:update_closest_threshold}).

\subsection{Performance Evaluation}

We now present the performance evaluation results of our DOF-ID architecture, where we set $c = 0.75$ and $\Theta = 0.65$. In addition, we used the DRNN model that has $10$ neurons in each cluster and has the following parameter settings: $p = 0.05$, $r = 0.001$, and $\lambda^+ = \lambda^- = 0.1$.

\begin{figure}[h!]
	\centering
    \includegraphics[scale=0.25]{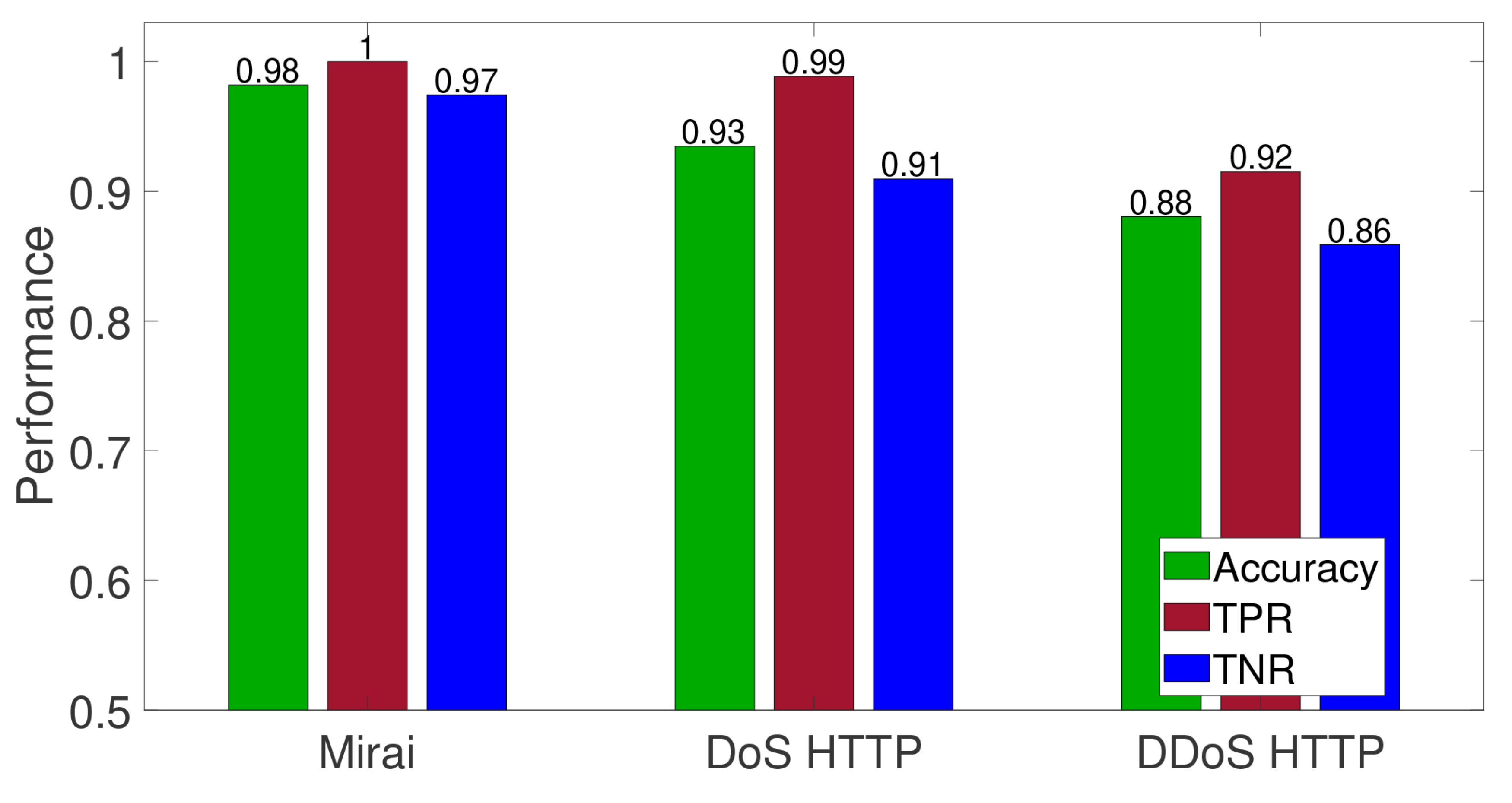}
    \caption{Performance of the DOF-ID architecture for each node among Mirai, DoS HTTP, and DDoS HTTP with respect to Accuracy, TPR, and TNR}
    \label{fig:performance_DOF-ID}
\end{figure}

Figure~\ref{fig:performance_DOF-ID} displays the average performance of DOF-ID with respect to Accuracy, True Positive Rate (TPR), and True Negative Rate (TNR). 
The results in this figure show that each node (i.e. Mirai, DoS HTTP and DDoS HTTP) achieves above $0.86$ detection performance with respect to all metrics. One may also see that although the nodes suffer from some false positive alarms (shown by the TNR metric), all nodes detect local intrusions with a considerably high performance (shown by the TPR metric).  

\begin{figure}[h!]
    \centering
    \includegraphics[scale=0.25]{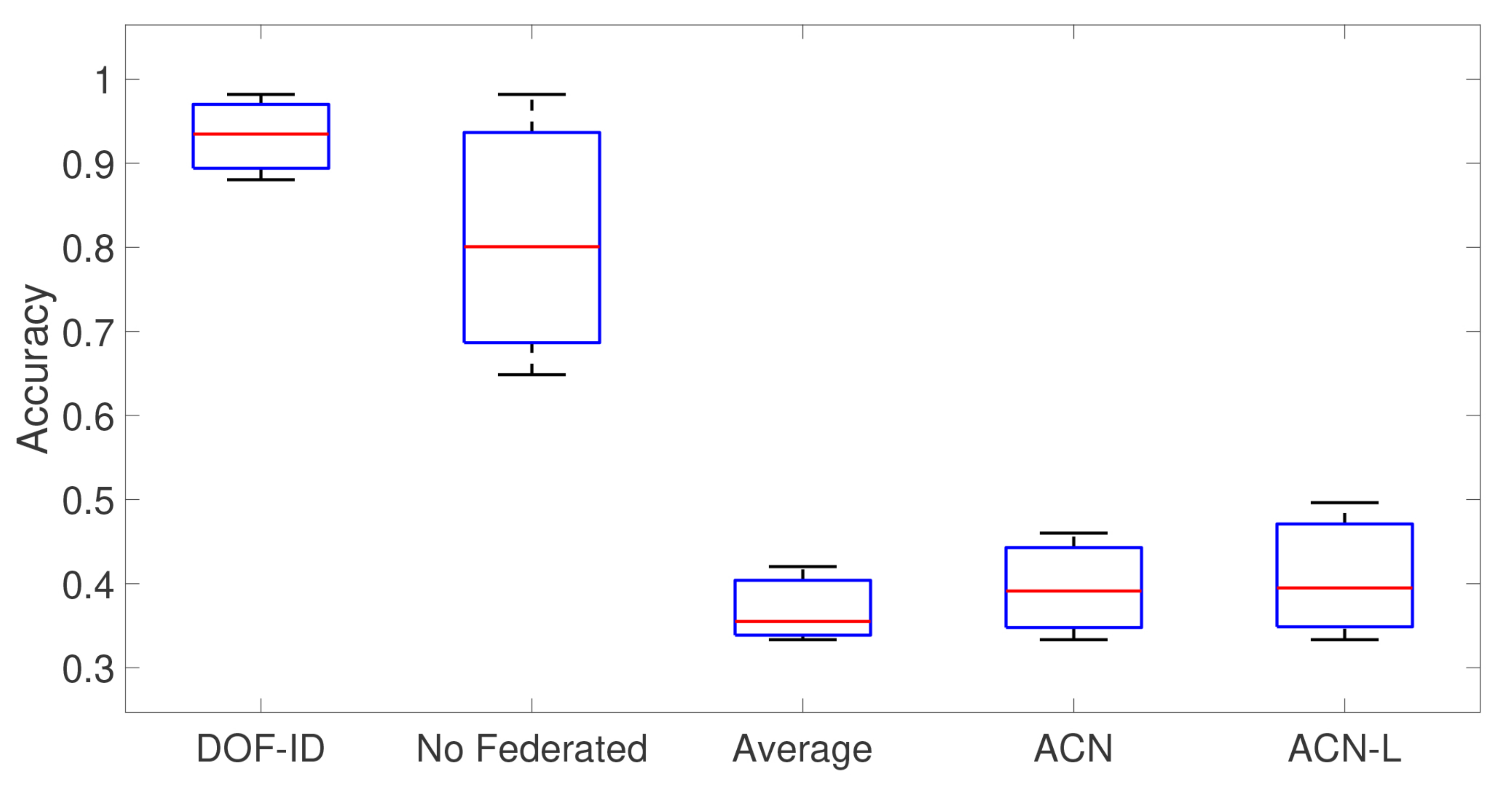}

    \includegraphics[scale=0.25]{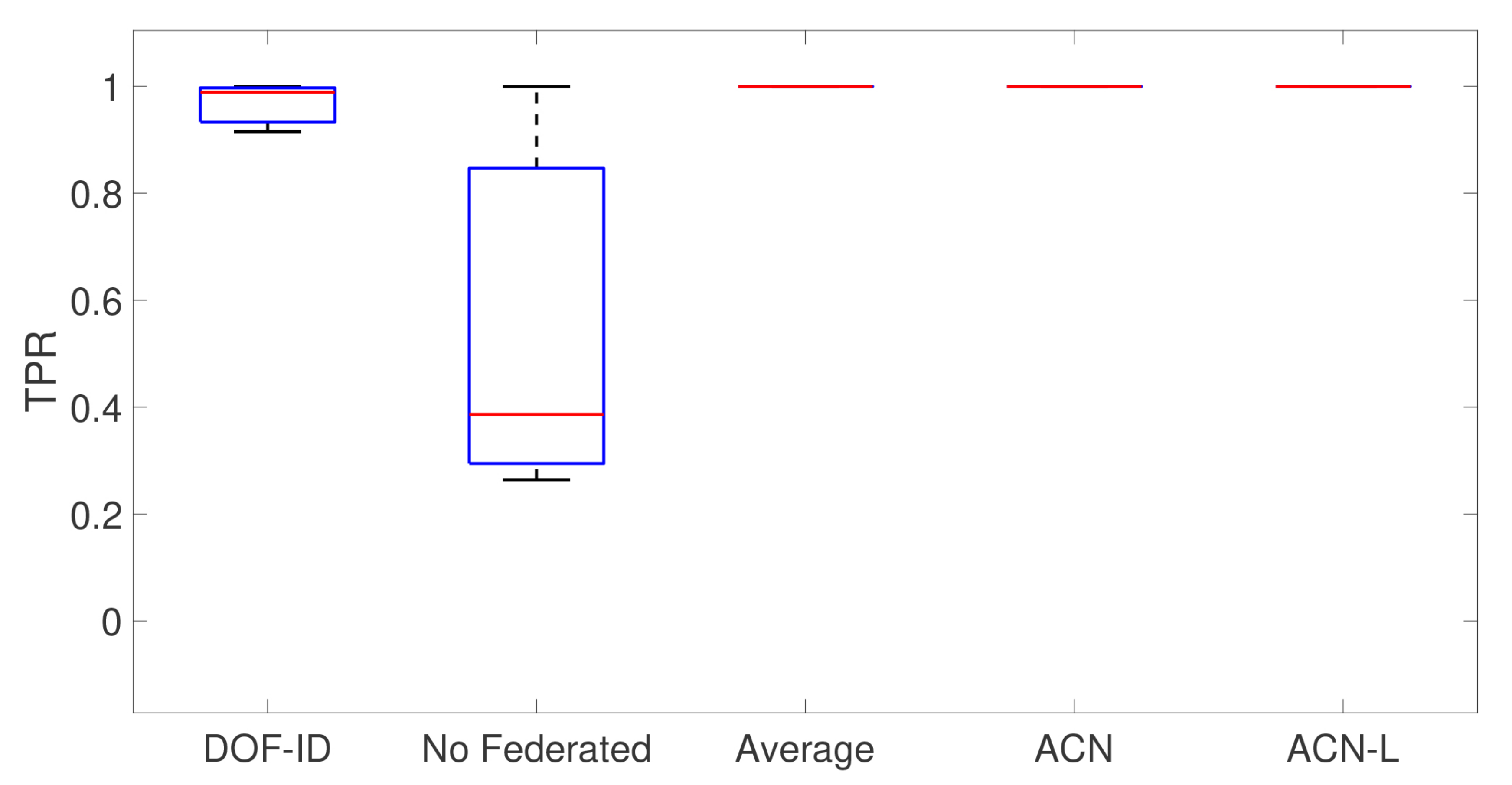}

    \includegraphics[scale=0.25]{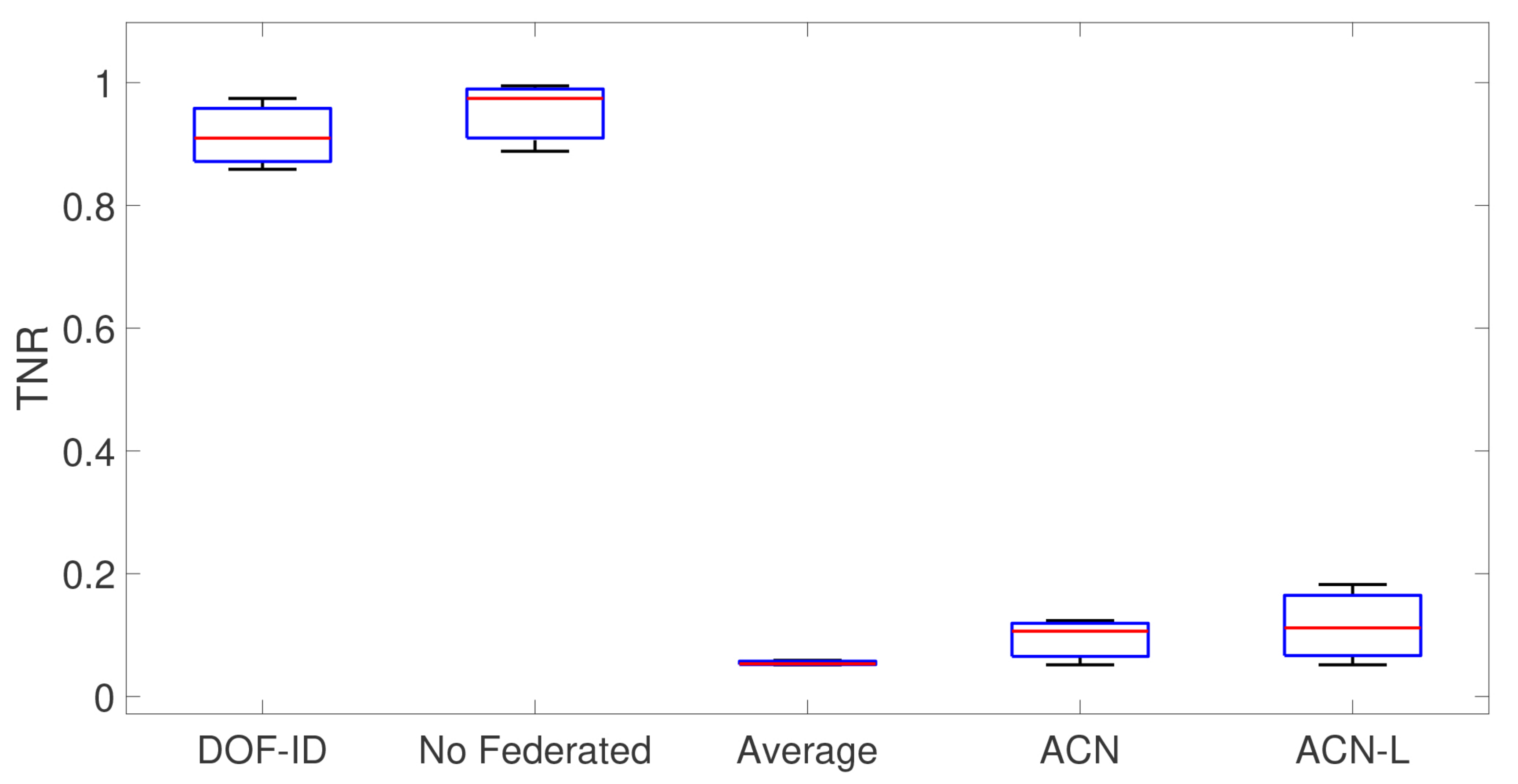}
    \caption{Performance comparison of DOF-ID with benchmark methods (No Federated, Average, ACN, and ACN-L) with respect to accuracy (top), TPR (middle), and TNR (bottom) presented as a box-plot across all nodes, namely Mirai, DoS HTTP and DDoS HTTP}
    \label{fig:performance_comparison}
\end{figure}

We also compare the performance of DOF-ID with benchmark methods in Figure~\ref{fig:performance_comparison}. The results in Figure~\ref{fig:performance_comparison} (top) show that the proposed method has the best accuracy among all methods compared. Another important observation of this figure is the poor performance of the averaging over all collaborating nodes. This is an expected result as network traffic across nodes varies considerably.

The evaluation results further show that FL-based methods (i.e. DOF-ID, Average, ACN, and ACN-L) significantly improve the detection performance measured by TPR in Figure~\ref{fig:performance_comparison} (middle), while they mostly tend to raise more false positive alarms compared to local learning as shown in Figure~\ref{fig:performance_comparison} (bottom). On the other hand, the proposed DOF-ID method appears to have a small decrease in TNR (i.e. a slight increase in false alarms) but a significant improvement in the detection rate, TPR.


Finally, we measure the training time of the proposed and existing methods, including the time required for federated update since the local learning time is the same (with negligible random deviations) for all models, i.e., $19.2~ms$ on average. However these measurements do not take the system level interactions into account when several servers communicate, nor do we measure the substantial overhead on a system when it comes under attack.
The measurements reported in Table~\ref{tab:update_time} show that the time spent by DOF-ID in addition to local learning is about $30~ms$ for each node. This time is significantly larger than other methods as its operations are more advanced and detailed. Also, a method is considered acceptable for a real-time application if the total operation time spent on local and federated learning and detection is shorter than the window length $T_n$. For the DOF-ID architecture, the total operation time is $48.91~ms$ on average, including learning time of $19.2~ms$, federated learning time of $29.6~ms$, and detection time of $0.11~ms$.

\begin{table}[]
\centering
\renewcommand{\arraystretch}{1.5}
\normalsize
\caption{Average time spent in microseconds ($\mu s$) by each method for the federated update in a single window}
\begin{tabular}{|c|c|c|c|}
\hline
& \textbf{Mirai}  & \textbf{DoS HTTP} & \textbf{DDoS HTTP} \\ \hline
\textbf{DOF-ID}  & $29.6 \times 10^3$       & $29.7  \times 10^3$          & $29.7  \times 10^3 $          \\ \hline
\textbf{Average} & $34.3$           & $33.3$              & $32.8$               \\ \hline
\textbf{ACN}     & $74.5$           & $66.4$              & $66.3$               \\ \hline
\textbf{ACN-L}   & $106.8$          & $101.1$             & $113.2$              \\ \hline
\end{tabular}
\label{tab:update_time}
\end{table}

\section{Conclusions}\label{sec:Conclusion}

This paper considers the IDS needs od distributed systems, such as Supply Chains, and proposex a novel Decentralized and Online Federated Learning Intrusion Detection (DOF-ID) architecture to improve the detection performance of anomaly-based IDS using a DRNN model and SWBC decision maker, both of which learn using only normal “benign” network traffic. The presented DOF-ID architecture provides a collaborative learning system that enables each node to learn from the experiences of other collaborating nodes without violating data confidentiality. In this way, DOF-ID improves both local and global security levels of all collaborating nodes simultaneously, quickly and effectively eliminating the requirement for a large learning data. 

This paper also evaluates the performance of DOF-ID and compares it against the benchmark methods using two public well-known datasets, Kitsune and Bot-IoT. During the performance evaluation, the impact of FL on intrusion detection performance is also investigated. Our experimental results revealed that the proposed DOF-ID method significantly improves the detection performance with a small increase in false positive alarms compared to the same IDS structure learning only from local traffic. In addition, the proposed method has significantly superior performance (at least $15\%$ accuracy difference) over benchmark methods with higher computation time.

Future work will expand the experimental setup and evaluate the performance of the proposed DOF-ID architecture for large networked systems such as supply chains, smart grids or large IoT networks. In this context, it wll be interesting to evaluate how these Federated IDS architectures may affect the overall energy consumption of the systems where they are installed \cite{EPN,Pernici,Kuaban23}. It would also be interesting to address the distributed communication costs, and the possible performance slowdown that may be caused by federated learning, as well as the new vulnerabilities that may be introduced to the learning process itself in the  proposed DOF-ID architecture. 

Accordingly, we shall analyse the time, bandwidth and energy requirements of this architecture due to P2P parameter exchange and investigate the security breaches that may aim to leak or corrupt IDS parameters during their transfer, as well as the possibility that the 
FL process itself may come under attack in a distributed system of systems \cite{X}.

\section*{Acknowledgements}

This research has been partially supported by the European Commission Horizon Europe, the Framework Programme for Research and Innovation (2021-2027), as part of the DOSS Project under Grant Agreement No: 101120270.
\bibliographystyle{IEEEtran}
\bibliography{references_federated,datasets,RNN,IDS}
\end{document}